\documentclass[aps,prl,twocolumn,tightenlines,superscriptaddress]{revtex4-1}
\usepackage{epsfig,rotating}
\usepackage{multirow}
\usepackage{amsbsy}
\usepackage{stmaryrd} 
\usepackage{braket}
\usepackage{amsmath}

\begin{document}
\title{The semi-quantum game of life}

\author{D.\ A.\ Faux}
     \affiliation{Department of Physics, University of Surrey,
       Guildford, Surrey GU2 7XH, United Kingdom}
    \author{M.\  Shah}
\author{C.\  Knapp}

   \date{\today}

\begin{abstract}
Conway's classic game of life is a two-dimensional cellular automaton in which each cell, either alive or dead, evolves according to rules based on its local environment. The semi-quantum game of life (SQGOL) is an adaptation in which each cell is in a superposed state of both dead and alive and evolves according to modified rules. Computer simulation of the SQGOL reveals remarkable complexity and previously unseen game behaviors.  Systems evolve to a ``quantum cloud" with a liveness distribution of mean $<\!a\!>=0.3480\pm0.0001$ and standard deviation $\sigma =0.0071$ which is dependent solely on the evolutionary rules.  Transient lifeforms emerge from the cloud. Semi-quantum still-lifes are discovered including the qutub  which contains 4 live cells and 4 semi-quantum cells. A solitary qutub  placed in an otherwise empty universe may act as a seed to reproduce child qutubs, one or more classical and/or semi-quantum lifeforms, oscillators, a quantum cloud or death depending on the initial state.  Evolution to the quantum cloud occurs chaotically with floating-point errors providing the butterfly effect.  Evolutionary outcomes scale in a fractal-like manner.  
\end{abstract}

\maketitle

John von Neumann and Oskar Morgenstern are credited with introducing game theory in 1944 \cite{VonNeumann.1928,von.1944} prompting applications in economics, business,  and the sciences.  
Probably the most well-known game is Conway's classic game of life \cite{gardner_1970} (CGOL) based on a two-dimensional (2D) cellular automaton where each cell is either 0 (dead) or 1 (alive). The system evolves in discrete time steps according to a set of simple rules  \cite{gardner_1970} which depend only on the number of live cells contained in its 8 surrounding cells, known as the Moore neighborhood \cite{ceccherini-silberstein_coornaert_2013}. The game therefore transmits information one cell at a time and is deterministic, its evolution depends solely on the initial state.   The rules are applied simultaneously to each cell at each time step, called a generation \cite{gardner_1970}, so that the game is homogeneous in space and time.  The rules established by Conway seek to mimic the success or demise of life based on environment; if the number of live cells in the Moore neighborhood is too high or too low the cell dies due to overcrowding or loneliness but flourishes if the surrounding liveness is optimum.  In some senses the CGOL mimics populations of viruses or bacteria whose fortune depend on their immediate surroundings. Conway's GOL is interesting because complex behaviors emerge from simple rules.  Many multi-cellular lifeforms have been identified which may be classed as ``still lifes" (static or unchanging), ``oscillators" which return to their original state after a number of generations, ``spaceships" which move, plus other more exotic species.
 
The quantisation of cellular automata was recognised as important in the development of quantum computation following the suggestion by Feynman in 1982 \cite{feynman.1982}.
The quantisation of mathematical games eventually followed \cite{eisert.1999}.  The semi-quantisation of Conway's classic game in 2008 by Flitney and Abbott \cite{flitney_abbott_2008} (SQGOL) used non-time-reversible rules that recovered the CGOL for purely live and dead cells.  A fully-quantum game of life followed in 2012 \cite{bleh.2012} with adapted and time-reversible evolutionary rules hinting at complexity in one dimensional simulations similar to the CGOL.  The obvious relevance to quantum computation prompted the study of 3D GOL analogues within the computer science community and the emergence of behaviors shown to mimic signals, wires and gates  \cite{arrighi.2012intrinsically,arrighi.2012partitioned}.

The SQGOL by contrast has received scant attention despite its potential for rich and complex life-like behavior.   The evolution of each cell is now dependent on quantum mechanical operators acting upon each cell in the system applied simultaneously at each generation \cite{flitney_abbott_2008}.  
Each cell is described by a normalised state $\ket{\psi}$ which is a superposition of alive $\ket{1}$ and dead $\ket{0}$ states such that \\[-3mm]
\begin{equation}
\ket{\psi} = a \ket{1} + b \ket{0} = a\begin{pmatrix} 1 \\ 0 \end{pmatrix} + b \begin{pmatrix} 0 \\ 1 \end{pmatrix} = \begin{pmatrix} a \\ b \end{pmatrix}. \label{eq:1}
\end{equation}
Here $a$ and $b$ are positive real constants. The evolutionary rules are based on the operators of ``birth" ($\hat{B}$), ``death" ($\hat{D}$), and ``survival" ($\hat{S}$) given by, 
\begin{eqnarray}
\hat{B} = \begin{pmatrix} 1 & 1 \\ 0 & 0 \end{pmatrix} \mbox{~~~}
\hat{D} = \begin{pmatrix} 0 & 0 \\ 1 & 1 \end{pmatrix} \mbox{~~~}
\hat{S} = \begin{pmatrix} 1 & 0 \\ 0 & 1 \end{pmatrix}. \label{eq:2}
\end{eqnarray}
The application of each operator also includes a normalisation step so that \\[-3mm]
\begin{equation}
\hat{B} \ket{\psi} = \ket{1} \mbox{~~~~} \hat{D} \ket{\psi} = \ket{0}
\mbox{~~~~} \hat{S} \ket{\psi} = \ket{\psi}. \label{eq:3}
\end{equation}
The SQGOL evolves each cell according to a set of rules which depend on the  liveness, $A$, of the Moore neighborhood defined as \cite{flitney_abbott_2008}
\begin{equation} 
A = \sum_{ik=1}^{8} a_k \label{eq:4},
\end{equation}
where $a_k$ is the liveness of the $k^{\rm th}$ cell in the Moore neighborhood.  The next generation is obtained using the operation 
\begin{equation}
\begin{pmatrix} a' \\  b' \end{pmatrix} = \hat{G} \begin{pmatrix} a \\ b \end{pmatrix} \label{eq:5}
\end{equation}
where $\hat{G}$ is provided in Table \ref{table1} \cite{flitney_abbott_2008}.  The SQGOL rules reproduce the laws of the Conway's CGOL in the cases where $A$ is an integer value. 

\begin{table}
	\centering
	\begin{tabular}{c  c} \hline\hline
		\textbf{~~~~~~~~~A~~~~~~~~} & \hspace{2.5cm}\textbf{$\hat{G}$}\hspace{2.5cm} \\
		\hline
		$0 \leq A \leq 1$ & $\hat{D}$ \\
		$1 < A \leq 2$ & $(\sqrt{2}+1)(2-A)\hat{D} + (A-1) \hat{S}$ \\
		$2 < A \leq 3$ & $(\sqrt{2}+1)(3-A)\hat{S} + (A-2) \hat{B}$ \\
		$3 < A \leq 4$ & $(\sqrt{2}+1)(4-A)\hat{B} + (A-3) \hat{D}$ \\
		$A \geq 4$ & $\hat{D}$ \\ \hline\hline
	\end{tabular}
	\caption{$\hat{G}$ operator for different values of $A$}
	\label{table1}
\end{table}

A cell $\ket{\psi}$ is therefore a two-level system, or qubit, which interacts with its immediate environment comprising the 8 cells of the Moore neighborhood. The environment interacts with the cell leaving its state $\ket{\psi}$ unchanged, collapsing the state to  $\ket{1}$ or $\ket{0}$, or changing the state to a new state $\ket{\psi^\prime}$. The evolution is not time-reversible and has similarity with the weak-measurement process in which a particle's state interacts, or exchanges information, with its local environment.  Unlike the weak measurement, the extraordinary emergent complexity of the SQGOL, which we now demonstrate, arises because {\em each} cell simultaneously interacts with and is influenced by its neighborhood at each generation.

The SQGOL was executed on a $100 \times 100$ grid with periodic boundary conditions.  To initiate the system, a fraction $f$ of cells was assigned a liveness $a$ chosen via a random number uniformly distributed between 0 and 1. The value of $b$ for each cell is obtained by normalisation.  The remaining cells are assigned as dead, that is $a=0, b=1$.   The system is then evolved according to the rules presented as Eqs.~\ref{eq:1}--\ref{eq:5} and Table~\ref{table1}.

\begin{figure}[tbh!]			
	\centering
	\includegraphics[width=10. cm, height=6.0cm, trim={4cm 3.5cm 3cm 1cm},clip]{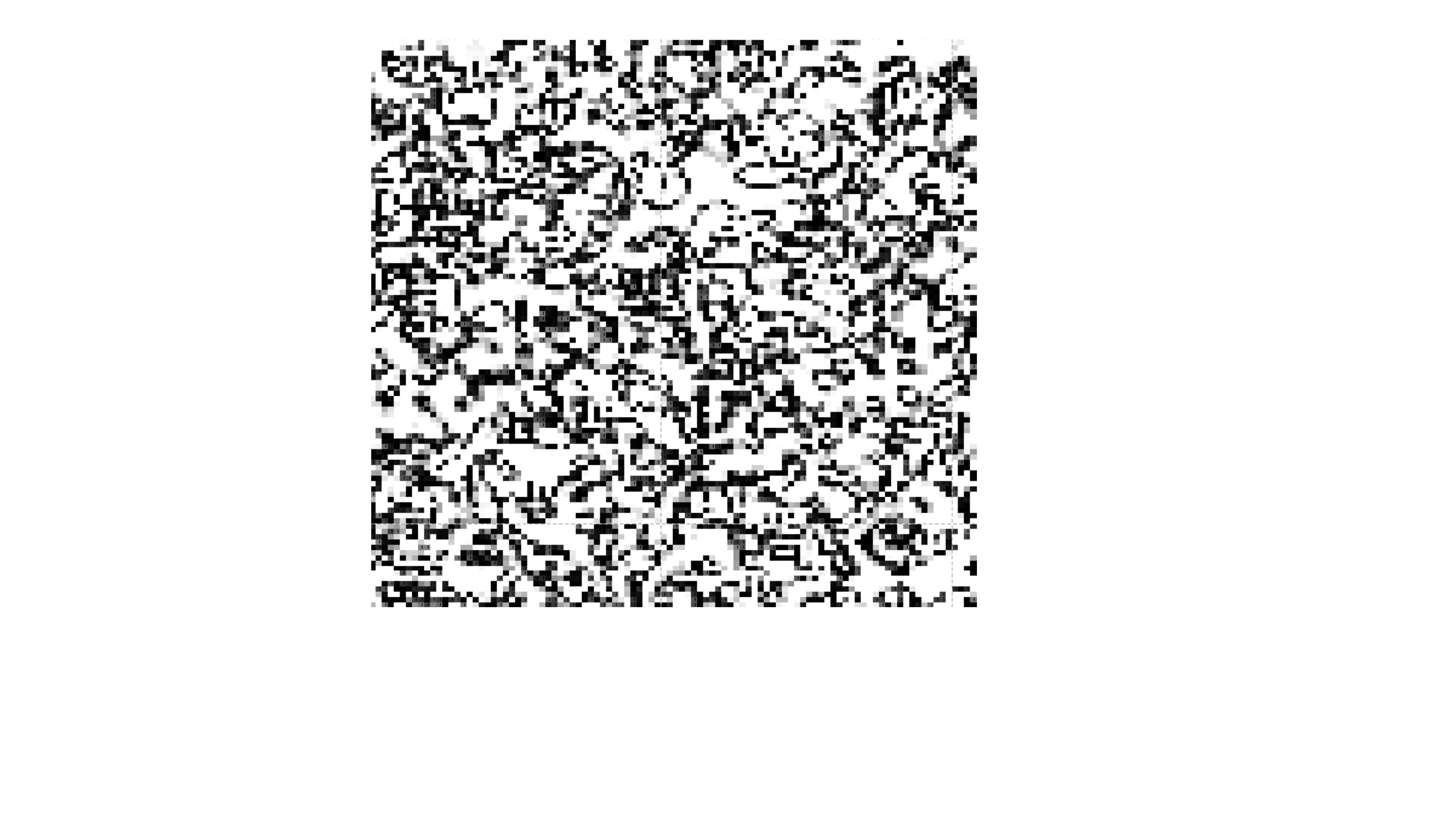}\\[-5 mm]
	\caption{The cell liveness parameter $a$ is plotted as dead (black), live (white) and semi-live (gray scale) cells for a $100 \times 100$ system. }
	\label{Fig1_slime}
\end{figure}

The system reaches a dynamic equilibrium comprising dead cells, live cells and cells in superposed states referred to as ``semi-live" as shown in Fig.~\ref{Fig1_slime}.  The SQGOL steady state contrasts starkly with the CGOL which evolves to produce a variety of still-lifes and oscillators.  The liveness distribution in Fig.~\ref{Fig1_slime} is referred to here as a ``quantum cloud".  The square of the liveness, $a^2$, represents the liveness probability density function.

\begin{figure}[tbh!]			
	\centering
	\includegraphics[width=0.45\textwidth, trim={0.5cm 1cm 0.5cm 1cm},clip]{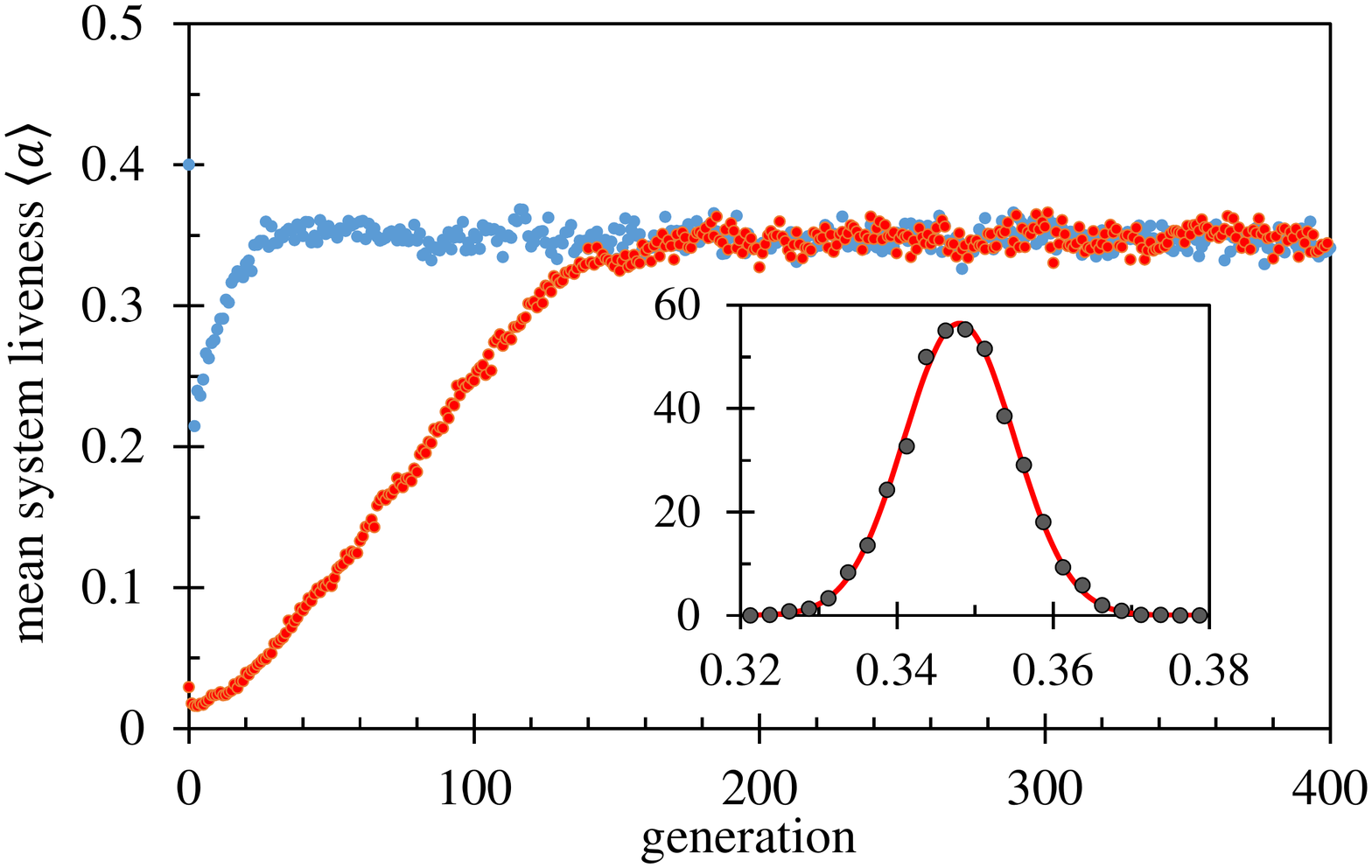}\\[-5 mm]
	\caption{The mean cell liveness $< \! a \! >$ of a SQGOL system of $100 \times 100$ cells with a starting fraction of live cells $f=0.2$ (red) and $f=0.8$ (blue) is presented. The inset demonstrates the Gaussian distribution of mean liveness.  }
	\label{Fig2_eqm}
\end{figure}

The evolution of the mean cell liveness $< \! a \! >$ is presented in Fig.~\ref{Fig2_eqm} for initial densities $f=0.2$ and $f=0.8$.  If $f$ is too small, the system dies out. If the system survives, a steady state dynamic equilibrium is established with system liveness $< \! a \! >$ of 0.3480$\pm$0.0001 to one standard deviation. The distribution of mean liveness  is a Gaussian with a standard deviation 0.0071 as shown in Fig.~\ref{Fig2_eqm}. The mean and standard deviation are ``universal constants" in the sense that they are dependent solely on the underlying evolutionary rules in table \ref{table1}.

\begin{figure}[tbh!]			
	\centering
	\includegraphics[width=8.5 cm, trim={0.5cm 1cm 0.5cm 2.5cm},clip]{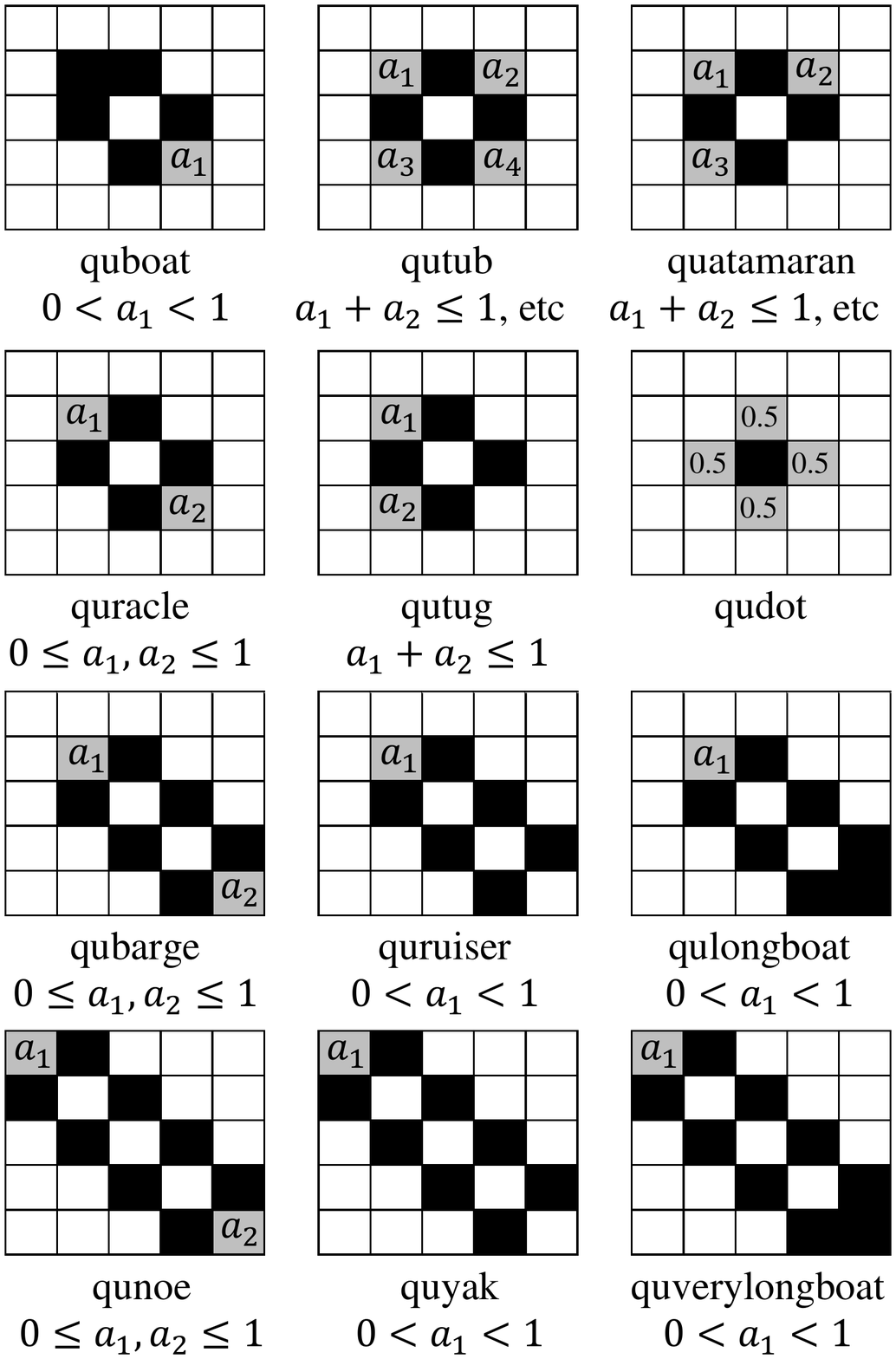}\\[-5 mm]
	\caption{12 quantum still-lifes are presented with the lifeness criteria necessary for longevity. \\[-10mm] }
	\label{Fig3_lifeforms}
\end{figure}

Unlike the CGOL, the SQGOL appears to produce no multi-cellular lifeforms.  However, close examination of evolving systems identifies transient lifeforms, existing perhaps for a small number of generations before encroachment and destruction by neighboring live or semi-live cells.  
Some {\em quantum} lifeforms have been identified and are included in Fig.~\ref{Fig3_lifeforms}.  Their names are empathetic to their CGOL equivalents where possible and others are provided new nautical labels.  The qutub and the quracle (named after the traditional Welsh wicker boat) are the most commonly observed transient lifeforms and some lifeforms are constructed.  Each structure is presented with the range of cell livenesses that provide still-lifes, that is they are invariant to the evolutionary rules. The structural pattern producing still lifes is evident from Fig.~\ref{Fig3_lifeforms}. The qudot only survives indefinitely with the precise livenesses shown.

The qutub is of special interest.  The qutub comprises $3\times 3$ cells of which 4 are live, 4 are semi-live and the central cell is dead.  The lifeform is unchanging if surrounded by dead cells (as shown in the figure) and if each semi-live cell satisfies the criterion that the livenesses $a_i$ of any pair of next-nearest (non-diagonal) neighbors sum to 1 or less.  The qutub is dormant, or hibernating, until an event occurs which changes one or more of the semi-live cells so that the stability criteria is no longer satisfied.  Now the qutub acts as a seed to spawn a variety of structures depending on the livenesses $a_i$ of the four semi-live cells.  

\begin{figure}[tbh!]			
 	\centering
 	\includegraphics[width=8 cm, height=12.0cm, trim={1.0cm 2cm 1.5cm 2.0cm},clip]{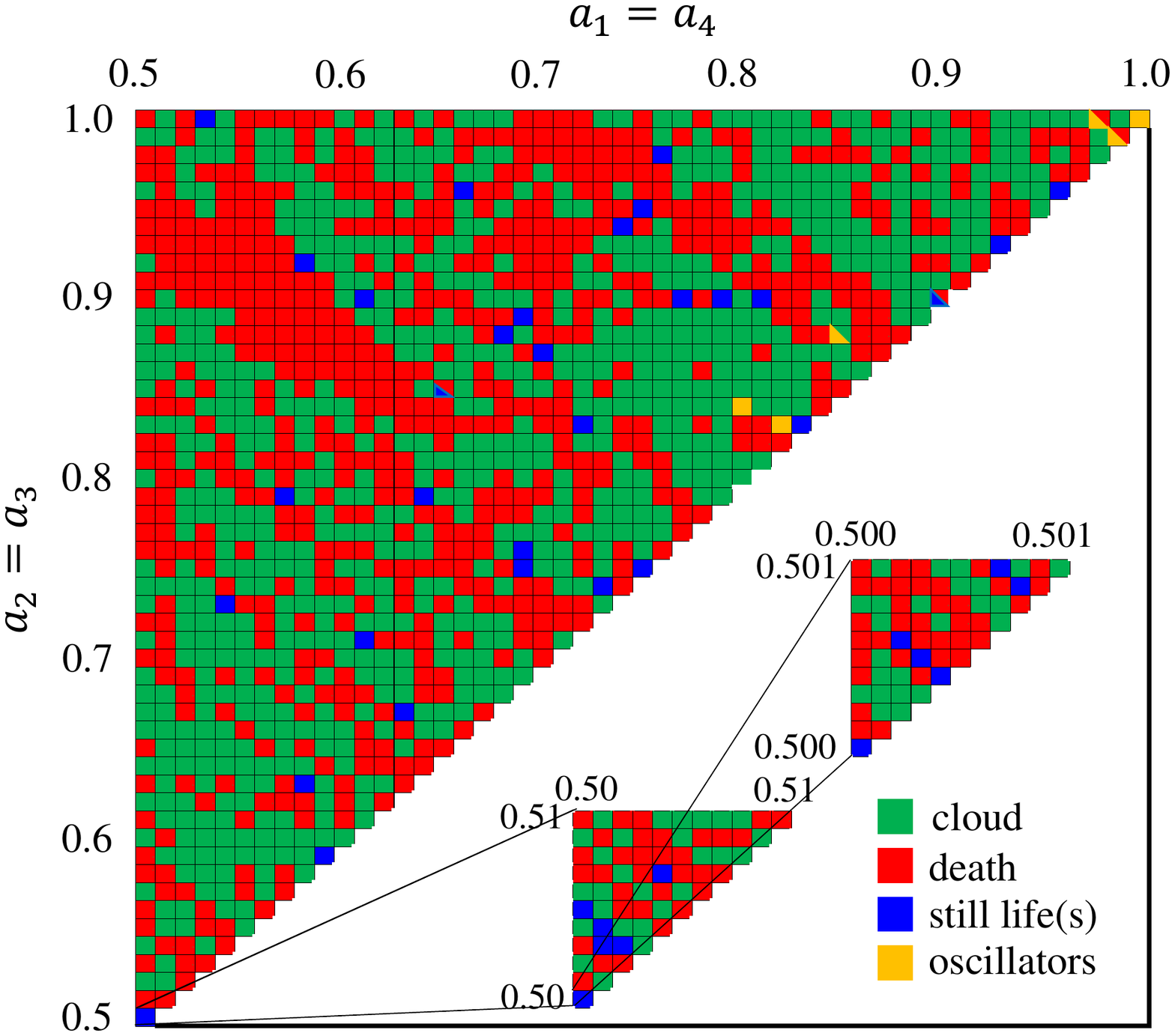}\\[-45 mm]
 	\caption{The evolutionary outcomes of a qutub seed with initial livenesses $a_i$ are presented. Half-cells indicate temporary forms. }
 	\label{Fig4_colorgrid}
 \end{figure}

The qutub structure in Fig.~\ref{Fig3_lifeforms} was set up as a seed in an otherwise empty universe of dimensions $100 \times 100$.  To limit the number of combinations, we set the liveness of the diagonal semi-live cells to be identical, that is $a_1=a_4$ and $a_2 = a_3$.  The evolution of the qutub seed was then explored for $a_1=a_4=0.5 ... 1.0$ and $a_2=a_3=0.5 ... 1.0$ at intervals of 0.01 as shown in Fig.~\ref{Fig4_colorgrid}. The leading diagonal therefore represents $a_1=a_2=a_3=a_4$.

The evolutionary outcomes summarised in Fig.~\ref{Fig4_colorgrid} show weak structure with identifiable diagonal bands from top left to bottom right.  These diagonal bands  have similar total initial qutub liveness.  With certain parameters, the qutub evolves with two- or four-fold symmetry and then dies out. With other parameters, the quantum cloud emerges such as that illustrated in Fig.~\ref{Fig1_slime}. Occasionally one or more classical or semi-quantum lifeforms emerge. Some qutub seeds evolve to form new qutubs, quracles or combinations of these.  One case leads to 16 classical blinkers, others to the classical tub or combinations of classical and semi-quantum lifeforms.  Occasionally a structure is stable for more than 4 generations before evolving to cloud or death.  Where discovered, these are indicated as half squares in Fig.~\ref{Fig4_colorgrid}.

The fine detail of the qutub evolution reveals a further interesting feature.  If $a_1=a_2=a_3=a_4=0.5$, the qutub is a still-life.  This is represented by the blue square at bottom left in Figure~\ref{Fig4_colorgrid}. An exploration of the semi-liveness parameter space at finer divisions of 0.001 for the range 0.50..0.51 and divisions of 0.0001 for the range 0.500..0.501 are shown in Fig.~\ref{Fig4_colorgrid}.  These yield a similar distribution of cloud, death and still-life outcomes with hints of structure across the diagonal.  This scaling similarity is analogous to fractal behavior.

\begin{figure}[tbh!]			
	\centering
	\includegraphics[width=8 cm, height=12cm, trim={0cm 1cm 6cm 1.9cm},clip]{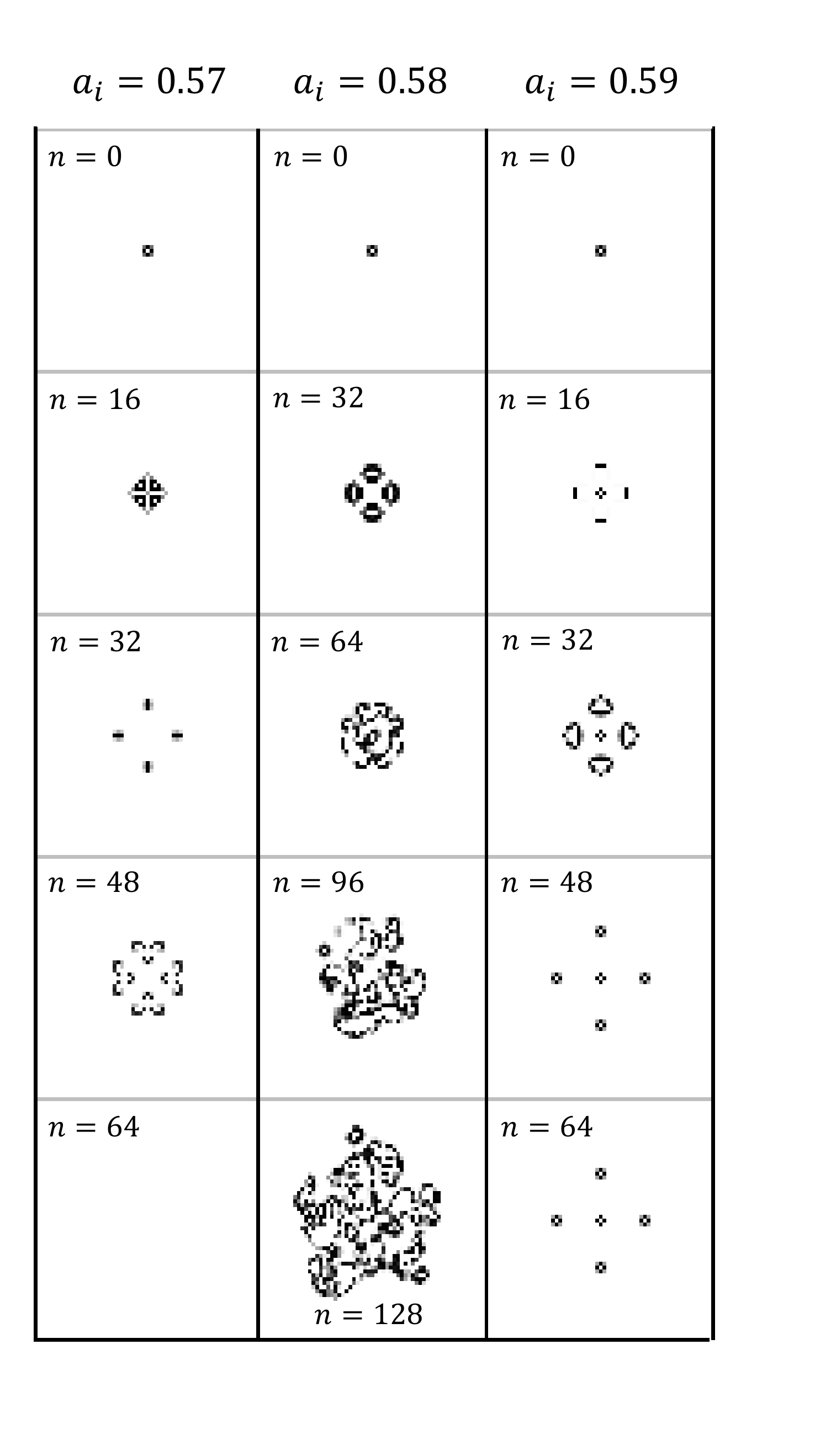}\\[-10 mm]
	\caption{The evolution of the qutub with $a_1=a_2=a_3=a_4$ equal to the values shown  at generation $n$ are presented. \\[-8mm]} 
	\label{Fig5_evolutions}
\end{figure}

The variety of outcome and sensitivity to the initial conditions is illustrated in Fig.~\ref{Fig5_evolutions} for three qutub seeds in which each $a_i=$0.57, 0.58 or 0.59. In the latter case the evolution produces a central tub plus four satellite qutubs.   Thus, the qutub  reproduces to provide 4 new still-life qutubs.  If the qutub seed has its four semi-live cells initiated at 0.57, the system evolves with four-fold symmetry but then dies out.  By contrast, at 0.58, the qutub seed initially evolves with four-fold symmetry as expected but the symmetry is suddenly lost and subsequent evolution produces a space-filling quantum cloud  as illustrated in Fig.~\ref{Fig5_evolutions}.  The initial symmetry is lost quite early in the evolutionary process and the emergent quantum cloud possesses exactly the same mean value and standard deviation as seen in Fig~\ref{Fig2_eqm}. Videos of each evolution are provided as supplementary material.

The loss of four-fold symmetry in the case of the qutub seed with $a_i=0.58$ is due to floating-point error. The SQGOL rules as listed in the table are different for different ranges of $A$ given in table \ref{table1}.  A specific calculation of $A$ may fall either side of a boundary due to floating-point error.  To take a simple example, a cell which has $A=3.000000$ might be calculated as 2.999997 and therefore be subject to a different rule.  The impact of floating-point errors in non-integer games is well known and does not feature in the integer-constructed CGOL.

Once a floating-point error causes a system asymmetry, the impact is surprisingly quick.  The results presented in all figures use double-precision floating point computation on a workstation.  The outcomes in Fig.~\ref{Fig4_colorgrid} were compared with single-precision computation on the same workstation and single-precision computation on a laptop executed with the same FORTRAN code.  The vast majority of final outcomes were identical in each case but a small number of lifeforms were found to be still-lifes with double-precision computation on the workstation but only temporarily stable with single-precision computation.  A small number of final outcomes were therefore completely different.  Interestingly, single-precision computation on the laptop produced fewer differences in evolutionary outcome when compared to the workstation using double-precision computation than for single- and double-precision computation on the {\em same} workstation.  The consequence is that the evolution of some systems develops chaotically with the floating-point errors providing the butterfly effect.

To the purist, the significant impact of floating-point error in the SQGOL is troublesome.  Our view is that this is a {\em strength} of the game.  Consider the SQGOL as a model of the evolution of lifeforms such as viruses, as claimed for the CGOL. We argue that the tiny fluctuations in computational precision due to floating point error are a reflection of the natural variation in response of lifeforms to the local environment in the evolutionary process. Viruses are not all perfectly identical and each virus will not in reality respond to its local environment in precisely the same manner to arbitrary precision.  The SQGOL provides the probability density function $a^2$ as a more compelling model for the evolution and spread of viruses than does the CGOL.

In summary, the semi-quantum game of life has been simulated for a 2D grid of cells providing rich and complex behaviour. The semi-quantum systems evolve to produce cells with a distribution of liveness $a$ with a mean $<\!a\!>=0.3480\pm0.0001$ and standard deviation $\sigma =0.0071$. Both $<\!a\!>$ and $\sigma$ are independent of initial configuration and system size and dependent solely on the evolutionary rules.  A set of semi-quantum still-lifes are discovered of which 12 are presented. The qutub, which contains 4 live cells and 4 semi-quantum cells, is shown to act as a seed when placed in an otherwise empty universe with semi-quantum cell livenesses outside the stability criterion.  The qutub evolves to produce child qutubs, one or more classical and/or semi-quantum lifeforms, the quantum cloud, oscillators, or dies out depending on the initial state.  The evolution to the cloud occurs chaotically due to processor-dependent floating-point errors.  The evolution of the qutub is shown to exhibit high sensitivity to the initial conditions and to exhibit fractal-like behaviour at reduced scales.  The SQGOL may provide a more physically-relevant simple game-based model for many real-life processes compared to its classical counterpart.

\bibliography{References_QGOL}

\end{document}